\documentclass{aa}

\usepackage{natbib}
\renewcommand\citep[1]{(\citealt{#1})}  
\newcommand\citepf[1]{(\citealt*{#1})}    

\newcommand\zzz[2]{#2}  

\usepackage{epsf}

\def\S{Sect.~}  

\def\k{\mbox{\rm\,km\,s$^{-1}$} }

\def\hMpc{\mbox{h$^{-1}$ Mpc}}

\def\hGyr{\mbox{h$^{-1}$ Gyr}}

\def\centreline{\centerline}
\def\.{{.}}
\def\gtapprox{\,\lower.6ex\hbox{$\buildrel >\over \sim$} \, }
\def\ltapprox{\,\lower.6ex\hbox{$\buildrel <\over \sim$} \, }
\def\sun{\odot}

\def\e{ {\scriptstyle \times} 10^}  

\def\arcs{\ifmmode {'' }\else $'' $\fi}     
\def\arcm{\ifmmode {' }\else $' $\fi}     
\def\deg{\ifmmode^\circ\else$^\circ$\fi}    
\def\ttimes{{\scriptstyle \times}}

\def\rinj{{r}_{\mbox{\rm \small inj}}}  
\def\dpm{{d_{\mbox{\rm \small pm}}}}   
\def\imax{{i_{\mbox{\rm \small max}}}}   
\def\imaxtiny{{i_{\mbox{\rm \tiny max}}}}   

\def\apj{ApJ}                 
\def\aanda{A\&A}            
\def\cqg{Class. Quant. Grav. }   %
\def\mnras{MNRAS}
\def\jaf{J. A. F.}  
\def\araa{ARA\&A}

\def\frtoday{le\space\number\day\space\ifcase\month\or
  janvier\or f\'evrier\or mars\or avril\or mai\or juin\or
  juillet\or ao\^ut\or septembre\or octobre\or novembre\or d\'ecembre\fi\space \number\year}

\newcommand\joref[5]{#1, #5, {#2 }{#3, } #4}

\newcommand\epref[3]{#1, #3, #2}

\def\fgeom{ 
\begin{figure} 
\centreline{\epsfxsize=8cm
\zzz{\epsfbox[80 0 542 502]{"`gunzip -c z0z1z2.ps.gz"} }
{\epsfbox[80 0 542 502]{"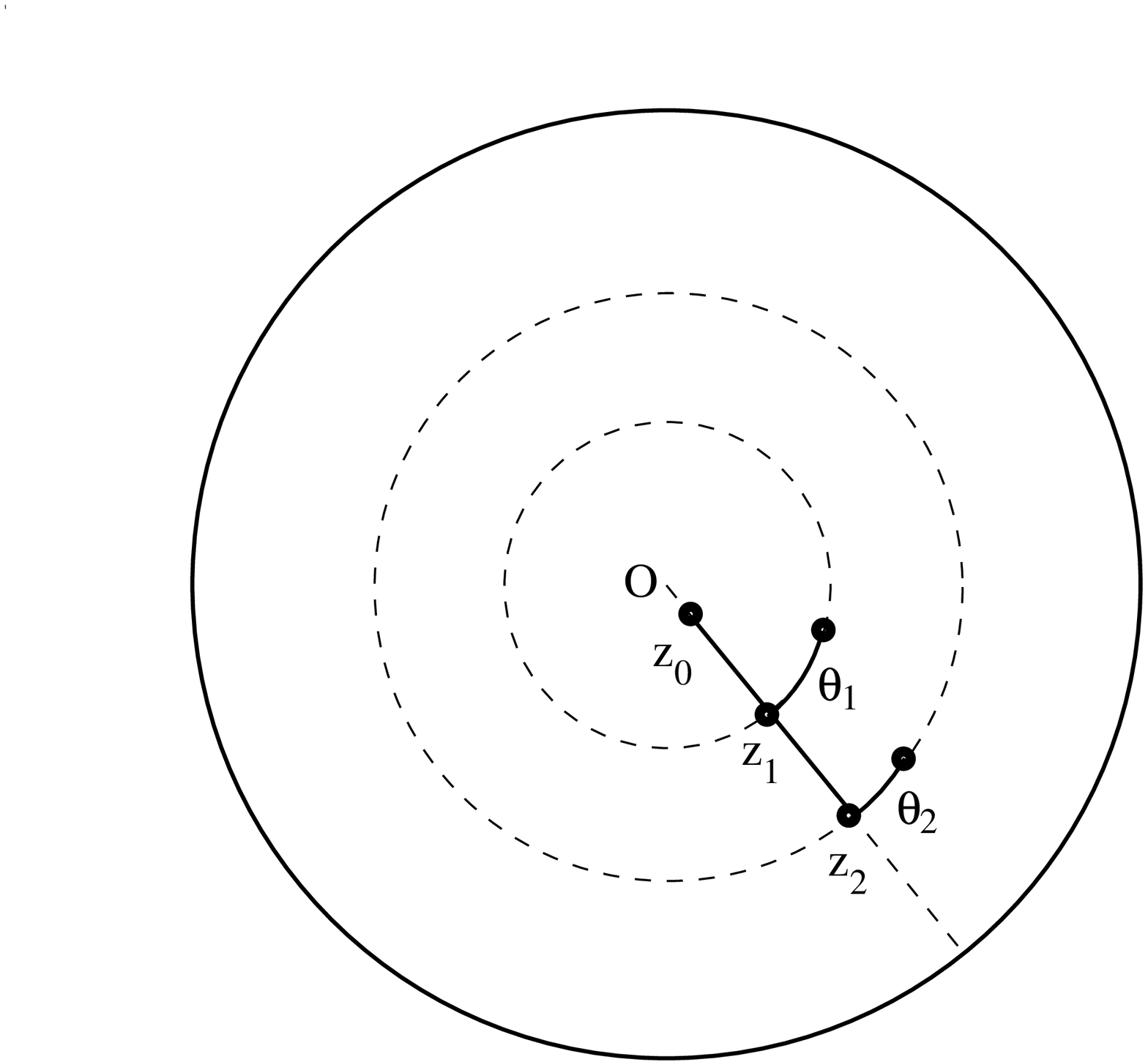"} }
}
\caption{\label{f-geom} 
Relative positions within the covering space 
of topological images used in the calculation, in spherical 
coordinates centred at the observer and limited by the
surface of last scattering. For the radial
component, three images are assumed to line up in an exactly
radial direction, at redshifts $z_0, z_1,$ $z_2,$ covering two equal
proper distance intervals, each of $2\protect\rinj$. For the tangential component, 
a pair of images at $z_1$ separated along a circular arc 
by $2\protect\rinj$ and another pair of images separated by $2\protect\rinj$ 
along a circular arc at $z_2$ are considered. This latter
is only a useful theoretical construct and would not (in general) 
occur in reality. 
If the curvature
is zero, then the angles subtended at $z_1$ are related by
$\theta_2/\theta_1 = d(z_1)/d(z_2),$ but for a non-zero curvature,
additional sinh or sin factors are required. The proportions 
shown in the figure are correct if the distance to $z_1$ is
one curvature radius, $\Omega_0\approx 0\.3, \lambda_0 \approx 0\.0$
and $2\protect\rinj \approx 3000$\hMpc.
}
\end{figure} 
}

\def\fconfirm{ 
\begin{figure} 
\centreline{\epsfxsize=8cm
\zzz{\epsfbox[ 0 0 393 164]{"`gunzip -c confirm.ps.gz"} }
{\epsfbox[ 0 0 393 164]{"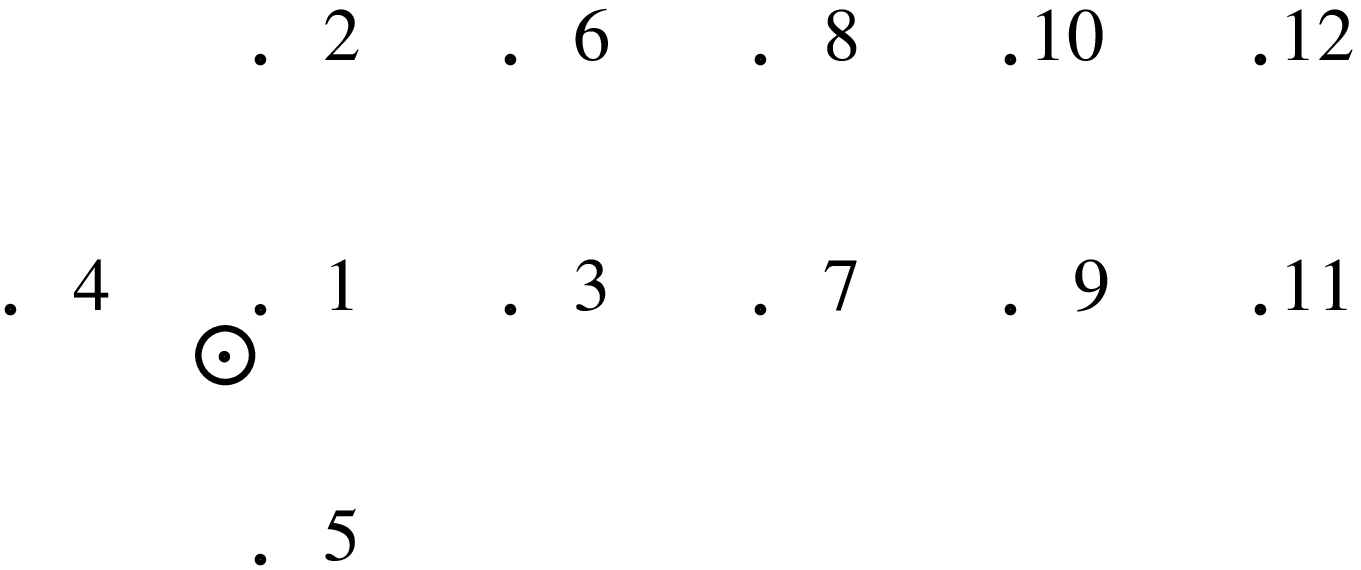"} }
}
\caption{\label{f-confirm} 
Simulated order of observationally 
confirming positions of topological images
of the hypothetically single cluster Coma/ 
RX~J1347.5-1145/ CL~09104+4109. The three known clusters are
numbered 1,2,3 respectively; following numbers indicate successive 
`observations'. The Sun is indicated near the position of the 
Coma image.
}
\end{figure} 
}

\def\fomBradi{ 
\begin{figure*} 
\centreline{\epsfxsize=8cm
\zzz{\epsfbox[41 23 459 452]{"`gunzip -c om_1000radi.ps.gz"} }
{\epsfbox[41 23 459 452]{"om_1000radi.ps"} }
\epsfxsize=8cm
\zzz{\epsfbox[41 23 459 452]{"`gunzip -c cc_1000radi.ps.gz"} }
{\epsfbox[41 23 459 452]{"cc_1000radi.ps"} }
}
\caption{\label{f-om_1000radi} 
Radial dependence of
uncertainty in $\Omega_0$ (left) 
and $\lambda_0$ (right) 
on redshifts of topological images, assuming
a cosmological redshift uncertainty of $\Delta z = 0\.002$ and
$2\protect\rinj=1000$\hMpc.  
The three curves for each set of
curvature parameters are for three topological images at $z_0 < z_1 <
z_2$ in a radial direction, such that $d(z_1)-d(z_0) = $ 
$d(z_2)-d(z_1) =$ $2\protect\rinj$.
}
\end{figure*} 
}

\def\fomAtang{ 
\begin{figure*} 
\centreline{\epsfxsize=8cm
\zzz{\epsfbox[41 23 459 452]{"`gunzip -c om_1000tang.ps.gz"} }
{\epsfbox[41 23 459 452]{"om_1000tang.ps"} }
\epsfxsize=8cm
\zzz{\epsfbox[41 23 459 452]{"`gunzip -c cc_1000tang.ps.gz"} }
{\epsfbox[41 23 459 452]{"cc_1000tang.ps"} }
}
\caption{\label{f-om_1000tang} 
Tangential dependence of uncertainty in 
$\Omega_0$ (left) and $\lambda_0$ (right) on redshifts
of topological images, assuming a cosmological redshift uncertainty of
$\Delta z = 0\.002$ and $2\rinj=1000$\hMpc.
The two curves for each set of
curvature parameters are the redshifts $z_1 <
z_2$ such that $d(z_2)-d(z_1) =$ $2\rinj$ and that there exists 
a pair of topological 
images separated by $2\rinj$ along an arc at $z_1$, and another pair
also separated by $2\rinj$ at $z_2$.
}
\end{figure*} 
}

\def\tsim{
\begin{table} 
\caption{\label{t-sim} Uncertainty in estimation of $\Omega_0$ 
for a simulation based on a candidate 3-manifold which is 
(hypothetically) confirmed by successive observations of 
cluster images at predicted positions $i=4,...,12.$ Uncertainties
of $\Delta z=0.002$ in the use of redshift as a cosmological distance 
indicator are used to randomly perturb the positions. 
The `best estimate' of the value of 
$\Omega_0$ is then deduced from the perturbed values. 
The input value used is $\Omega_0=0.3$. The model is flat, 
so $\lambda_0 \equiv 1-\Omega_0.$
As a function of the total number, $\imax,$ of topological images
identified, according to the scheme in Fig.~\protect\ref{f-confirm},
the highest redshift of those objects $z$ 
and the mean and standard deviation of the deduced values 
of $\Omega_0$ ($\left< \Omega_0 \right>,$ $\Delta \Omega_0$ 
respectively) 
are listed here.
}
$$\begin{array}{cccc}
\hline
\mbox{\rule[-1.8ex]{0cm}{5.0ex}}  
 \imax & z & \left< \Omega_0 \right> &  \Delta \Omega_0 \\ 
\hline 
\mbox{\rule[-0ex]{0cm}{3ex}}  
   5 &    0.45 &    0.2981 &     12.1\e{-3}  \\ 
   6 &    0.69 &    0.3003 &     5.4\e{-3}  \\  
   7 &     1.01 &    0.3002 &     2.6\e{-3}  \\ 
   8 &     1.23 &    0.3003 &     2.2\e{-3}  \\ 
   9 &     1.85 &    0.3002 &     1.5\e{-3}  \\  
   10 &     2.11 &    0.3001 &     1.3\e{-3}  \\ 
   11 &     3.24 &    0.3002 &     1.2\e{-3}  \\  
   12 &     3.63 &    0.3000 &     1.2\e{-3}  \\  
\hline 
\end{array} $$
\end{table}
} 

\begin{document}

   \thesaurus{12  
(12.03.3; 
12.03.4; 
12.04.3; 
11.03.1; 
11.17.3; 
03.13.5) 
} 
   \title{Constraining curvature parameters via topology}

   \author{B. F. Roukema\inst{1,2}  
          \and
	J.-P. Luminet\inst{3}}

   \offprints{B. F. Roukema}

   \institute{Inter-University Centre for Astronomy and Astrophysics, 
    Post Bag 4, Ganeshkhind, Pune, 411 007, India (boud@iucaa.ernet.in)
\and Observatoire de Strasbourg, 
  11 rue de l'Universit\'e, F-67000 Strasbourg, France
\and 
DARC, 
Observatoire de Paris-Meudon, 5 place Jules Janssen, F-92195 Meudon Cedex,
France (Jean-Pierre.Luminet@obspm.fr)}

\date{received 12 November 1998, accepted 26 March 1999. }

\authorrunning{B. F. Roukema \& J.-P. Luminet}
\titlerunning{Constraining Curvature via Topology}
   \maketitle

   \begin{abstract}
If the assumption that physical space has a trivial topology is dropped,
then the Universe may be described by a multiply connected
Friedmann-Lema\^{\i}tre model on a sub-horizon scale.
Specific candidates for the multiply connected space manifold have already
been suggested. How precisely would a significant
detection of multiple topological images of a single object,
or a region on the cosmic microwave background,
(due to photons arriving at the observer
by multiple paths which have crossed the Universe in different directions),
constrain the values of the curvature parameters
$\Omega_0$ and $\lambda_0$?

 The way that the constraints on $\Omega_0$ and $\lambda_0$ 
depend on the redshifts
of multiple topological images and on their radial
and tangential separations is presented and
calculated. The tangential separations give the tighter
constraints: multiple topological images of known types
of astrophysical objects at redshifts $z \ltapprox 3$
would imply values of $\Omega_0$ and $\lambda_0$ 
preciser than $\sim 1\%$ and $\sim 10\%$ respectively.
Cosmic microwave background `spots' identified with 
lower redshift objects by the Planck or MAP satellites
would provide similar precision. This method is purely
geometrical: no dynamical assumptions (such as the virial
theorem) are required and the constraints are independent
of the Hubble constant, $H_0.$


   \end{abstract}

%

\section{Introduction}

Is the Universe `open', flat or `closed'? This is a major 
question in observational cosmology, where the meaning is generally
intended to be, respectively: 
`Is the curvature of the Universe negative, zero or positive?'

The latter question is quantified as the measurement of 
the density parameter, $\Omega_0,$ and the (dimensionless) 
cosmological constant, $\lambda_0,$ which together determine
the curvature 
$\kappa_0 \equiv \Omega_0 + \lambda_0 -1.$~\footnote{Note that
$\Omega_0, \lambda_0$ and $\kappa_0$ correspond 
to $\Omega, \Omega_\Lambda$ and $-\Omega_R$ respectively 
in \protect\citeauthor{Peeb93}'~(1993) notation.} The value of
$\kappa_0$ is negative, zero or positive for negative, zero or
positive curvature respectively.
Recent observations from faint galaxy counts, gravitational lensing, 
type Ia supernovae 
\citepf{FYTY90,FortMD97,ChY97,SCP9812}
and from the cosmic microwave background (CMB) 
favour an approximately zero curvature, though what seems 
to be the presence of geodesic mixing in the COBE observations of the CMB
would require a Universe which has negative curvature \citep{GurTor97}.

However, if the words `open' and `closed' in the 
former question are interpreted to mean `infinite' and `finite' 
in spatial volume, respectively, and if the curvature is not positive, 
then answering the question requires knowing
not only the curvature of the Universe but also its topology.
In fact, the question should then be reworded into a double 
question: 
`Is the Universe positively curved (hence finite)? 
If not, then is it infinite (``open'') or finite (``closed'') in volume?'
A flat or negatively curved (hyperbolic) universe with a 
Friedmann-\-Lema\^{\i}tre-\-Robertson-\-Walker metric 
can be either infinite or finite, a common example of the latter 
being the flat hypertorus.

Moreover, measurement of the topology of the Universe is not only
likely to be necessary to answer the 
question under the latter interpretation of the
words `open' and `closed', but it would help to measure their
quantitative meanings under the former interpretation as 
descriptions of curvature. The precision attainable is at least
as good as that expected from the Planck and MAP satellites.

It is hoped that the global topology of the Universe can be either
detected or shown in an assumption-free way to be non-measurable
within the next decade.
Several new observational methods to constrain, detect and/or measure
the global topology of the Universe have been recently developed 
\citepf{LLL96,Rouk96,Corn96,Corn98b,RE97,ULL99a}. 
The applications of these methods 
to observational programmes of major ground-based telescope and satellite
projects already started or expected in the coming few years
(SDSS, \citealt{Love98}; XMM, \citealt{Arn96}; 
Planck Surveyor; MAP)
may well result in 
estimates of the topological parameters of the Universe,
if these are measurable within the horizon defined in practice by the
surface of last scattering (SLS).

Indeed, specific candidates for 
the 3-manifold of the spatial hypersurface projected to the
present epoch have already been proposed \citepf{RE97,BPS98}. 

It is therefore timely to examine how precisely a significant
detection of multiple topological images of a single object
or region on the CMB (due to photons arriving at the observer
by multiple paths which cross the Universe in different directions)
would constrain the values of
$\Omega_0$ and $\lambda_0$. This is the purpose of this paper.

Mathematically, the constraint on the curvature parameters is a
consequence of the rigidity theorem of Mostow \& Prasad
\citep{Mos73,Pra73}, which constrains the curvature radius $R_C$
(eq.~\ref{e-defR_C}), and for a fixed value of $R_C$ is a consequence
of the effect that $\lambda_0$ has on the redshift-distance relation.
To our knowledge, using topology to constrain curvature
was first clearly mentioned in a cosmological context by
\citet{BernShv80}.

\subsection{The principle of constraining curvature parameters 
($\Omega_0,$ $\lambda_0$) from multiple topological images}

Cosmic topology is briefly described in \S\ref{s-topol}. 

In the case of a detection considered significant, photons emitted 
by a single collapsed object 
or from a single region of recombination epoch plasma 
would
have crossed the whole Universe in different directions and
probably more than once, before arriving at the observer in less
than the age of the Universe. These would be considered to emanate
from multiple topological images of a single object (or region).

The geometrical relationship between the apparent three-dimensional 
positions of the different topological images in the covering space
(the na\"{\i}ve observer's space)
is related to the shape of the fundamental polyhedron 
(parallelepiped in the
case of the hypertorus). The linear transformations (isometries) 
between the multiple images, in the covering space, 
are integer linear combinations
of the generators of a group of isometries. For the rectilinear 
hypertorus with no `twists', the covering space is $R^3,$
the topological images of a single object form a rectangular
prismical lattice, 
the isometries are simply translations, and the generators 
are three mutually orthogonal vectors.

These isometries would hold for some particular values of $\Omega_0$ and
$\lambda_0.$
If these curvature parameters are varied from the values for which
the identification was discovered, then the 
distances between the identified images will no longer 
be related in the same proportions,
and will no longer (in general) 
remain integer linear combinations of a set of generators.
In other words, if the curvature parameters are varied, then the
objects which previously had been identified as multiple topological 
images can no longer be explained as such.

Another way of looking at this is to consider the predictive
power of candidate 3-manifolds. 
Given a candidate 3-manifold, the positions and redshifts of 
multiple topological images of
all known objects visible to large redshifts can be predicted.
If the candidate is correct, then observational confirmation
of the predicted images will strengthen the hypothesis. 
However, if the candidate is correct in a limited 
redshift range to a certain precision, {\em but
slightly wrong values for the curvature parameters are assumed,}
then the multiple topological images in other redshift 
ranges will not occur at the
predicted positions. The error will depend on the redshift
and on the errors in the curvature parameters.

At low redshifts, the dependence of the metric on curvature 
is weak, so the constraints on $\Omega_0$ and $\lambda_0$ would
be weak. For multiple topological images at high redshifts 
distributed in the three-dimensional
space inside of the SLS, the constraints would be much stronger, 
i.e. the uncertainties to within which $\Omega_0$ and $\lambda_0$ could
be estimated would be much smaller. 

On the other hand, a detection of multiple topological imaging 
of circular subsets of the COBE maps
(\citealt{Corn98b}), at 
the SLS, without confirmation from topological images 
inside of the SLS, would determine the sign of the curvature, 
and if this is non-zero, would fix the ratio $R_H/R_C$
by the rigidity theorem, 
where $R_H$ is the particle horizon radius\footnote{Strictly speaking, 
what is constrained involves 
the distance to the SLS, which is slightly closer to us than 
the particle horizon that would be calculated if 
the matter-dominated regime is extended beyond the regime 
where it is physically correct. In order for the calculation to 
be mathematically correct, the difference between the two would
have to be taken into account.} and $R_C$ is the 
curvature radius [eq.~(\ref{e-defR_C})].

Only a relation between $\Omega_0$ and $\lambda_0$ 
would be known ($\Omega_0+\lambda_0=1$ would be the relation 
for a 3-manifold with $R^3$ as the covering space), and extra
information would be needed to find their individual values.
This is because the SLS is essentially a spherically 
symmetric two-dimensional surface, i.e. a sphere, so that there
is radial freedom in the relation between redshift and distance
which would allow ranges of the values of the two 
curvature parameters which would leave the mappings on the
sphere unchanged.

A CMB topology detection would, of course, be followed up by the
search for sub-SLS topological images. If these were found at low
redshifts $z \ltapprox 0\.5,$ then the constraints on $\Omega_0$ and
$\lambda_0$ would most likely remain weak. If they were found at
higher redshifts, $z\sim 3,$ then the constraints would be stronger.
However, the limited size of observational catalogues at higher
redshifts, and the probable relative scarcity of astrophysically
stable, bright objects at these redshifts, makes the latter a task
much less straightforward than might be hoped for na\"{\i}vely.

The goal of this paper is to estimate how precisely
$\Omega_0$ and $\lambda_0$ can be estimated, in practice, if
sub-SLS multiple topological images are detected to high significance.
The present value of 
the Hubble parameter, $H_0,$ is only a scaling factor, so would
not be constrained directly by topology detection. 
Conversely, the uncertainty of $H_0$ does not (to first order) 
affect the constraints on the curvature parameters by this method.
However, independent
constraints on $\Omega_0 h^{-2}$ or $\Omega_0 h$, 
such as from nucleosynthesis or from large scale structure, 
would provide good estimates of $H_0$ if $\Omega_0$ were estimated
to high precision (e.g. $<$1\%).\footnote{The Hubble constant
is parametrised as $h\equiv H_0/100$km~s$^{-1}$~Mpc$^{-1}$.}
Some comments regarding identification of objects in the CMB,
e.g. cold spots, with their evolved, local counterparts are
also made.

\subsection{Structure of paper and conventions}

In \S\ref{s-method}, the reader 
is briefly reminded of some elements of the 
geometry of Friedmann-Lema\^{\i}tre universes (\S\ref{s-geom}), 
and the method of deriving constraints on curvature parameters 
is explained (\S\ref{s-uncert}).
The resulting precisions to which 
$\Omega_0$ and $\lambda_0$ can be constrained is 
presented in \S\ref{s-res}. 
Discussion and conclusions are presented in \S\ref{s-disc}
and \S\ref{s-conc}.

Proper distances, in comoving coordinates, as in eq.~(\ref{e-dprop}),
are used throughout. 

\section{Method} \label{s-method}

\subsection{The geometry of spatial sections (`space')}  \label{s-geom}

The application of general relativity to observations under
the assumptions of a Friedmann-Lema\^{\i}tre universe 
results in three-dimensional spatial sections which may have
non-zero curvature and/or non-trivial topology\footnote{non 
trivial $\pi_1$ homotopy group}
(e.g. \citealt{deSitt17,Fried24,Lemait58}). Both possibilities were 
suggested in the pre-relativistic epoch by 
\citet{Schw00}. 

\subsubsection{Cosmological topology} \label{s-topol}

It is often implicitly hypothesized
that the topology of the Universe is simply-connected. 
Since the density parameter is within an order of magnitude
of unity, the observable sphere (inside the particle horizon) 
would in that case be 
much smaller than the Universe by a factor of $10^N$ where
$N \gg 1$ [e.g. see \citet{Linde96} for discussion of inflation
for zero versus negative curvature]. If this were correct, then
observational cosmology would not be the study of the Universe,
but only of a microscopic part of the Universe expected to be
representative of the whole under some scenarios. This, of course,
only provides an aesthetic criterion in favour of an observably
`small' Universe.

However, the requirement to have some sort of theory of 
quantum gravity at the Planck epoch ($t \sim 10^{-43}s$),
and the fact that topological
evolution in numerous dimensions is common in theories of particle
physics, make a finite universe with a non-trivial topology 
(non trivial $\pi_1$ homotopy group) quite a theoretically 
likely possibility.


For general reviews on cosmological topology, 
see \citet{LaLu95,Lum98,Stark98,LR99}.

For theoretical 
work on physical explanations of the generation of topology at the
quantum epoch, see recent articles such as 
\citet{MadS97,Carl98,Ion98,DowS98,DowG98,Rosal98,eCF98}.

Recent observational methods of constraining and/or 
detecting candidates for the multiply connected manifold representing space
include (i) cosmological microwave background (CMB), i.e. essentially 
two-dimensional  
methods (\citealt*{Stev93,LevSS98}; \citealt{Corn98b,Weeks98,Rouk99}) and
(ii) three-dimensional analyses of the distribution of discrete
cosmic objects, such as galaxy clusters and quasars
\citep{Fag87,Fag96,Rouk96,LLL96,RE97,RB98,LLU98,ULL99a}.
As mentioned above, some specific candidates for 
some or several of the generators defining the 3-manifold 
exist \citep{RE97,BPS98}. 

Several authors of the CMB methods have claimed lower limits
to the size of the Universe
of around 40\% of the horizon diameter based on observational 
data from the COBE satellite, particularly for 
the case of zero curvature, but these claims remain controversial.
\citet{Corn98a} argue that the resolution of COBE is
insufficient for any constraints on topology, but more fundamentally,
the problem is one of self-consistency of assumptions and hypothesis.
Even though the hypothesis tested is that of a small universe,
the assumptions made are those which are expected theoretically for
a large universe, and which have only been observationally supported
under the assumption of simple-connectedness.

If the Universe is observably multiply connected, then the standard
version of inflation cannot quite be correct. Even if inflation is
retained to solve various problems, it would be surprising to expect
that it would have resulted in exactly gaussian perturbation
statistics with a Harrison-Zel'dovich scale-free spectrum {\em on the
scale of the fundamental domain}. Moreover, justifying the latter
properties from a COBE data analysis under the assumption of
simple-connectedness would not be valid.  In fact, the articles
claiming constraints from COBE data generally find that a small
percentage of their perturbation simulations based on such assumptions
are consistent with the COBE data, but do not examine what properties of
the perturbations are required for small universe models and observations
to be consistent.

Here, the reader is just reminded of a few aspects 
of cosmological topology. A simple two-dimensional analogy for
a space (surface) with a non-trivial topology is the 2-torus, $T^2,$ 
which can be imagined equivalently either as 
\begin{list}{(\arabic{enumi})}{\usecounter{enumi}}
\item a torus placed in Euclidean 3-space, $R^3,$ but given an
{\em intrinsic} metric such that the curvature is zero everywhere on 
the surface; or as
\item a rectangle whose opposite sides are identified; or as
\item an infinite grid of multiple copies of a single rectangle 
corresponding to multiple copies of a single physical space.
\end{list}
Case (3) corresponds to what is termed the
{\em `universal covering space'}, in this case $R^2.$ 
A single copy of the rectangle, as in cases (1) or (2), is 
termed the `fundamental domain', or `Dirichlet domain', 
or in the three-dimensional case, the {\em `fundamental polyhedron'}.
Mathematically,
the 2-torus can be derived from $R^2$ by the isometric transformations
which are simply translations in two perpendicular directions,
i.e. $T^2 = R^2/\Gamma$, where $\Gamma$ is the group generated
by the two translations.

The corresponding three-dimensional space is the 3-torus, $T^3,$ 
for which case (1) is difficult to imagine, case (2) involves 
identification of opposing faces of a rectangular prism, and
case (3) has $R^3$ as the covering space. 

This is only one
example. In general, the 3-manifolds of cosmological interest 
(i.e. with constant curvature) can 
be represented in terms of a fundamental polyhedron embedded 
in the appropriate covering space ($H^3,$ $R^3$ or $S^3$ for 
negative, zero or positive curvature respectively), of which
faces are identified in a certain way. The metric is 
that of the covering space. The isometries mapping
one copy of the fundamental polyhedron to another within 
the covering space are linear (integer) combinations of
the {\em generators} which form the holonomy group $\Gamma.$

If the fundamental polyhedron is `smaller' than the SLS,
then multiple topological images of astrophysical objects 
or of temperature fluctuations averaged over small regions of space
should be visible. The smallest size of a (spatial) geodesic 
linking a point to itself (i.e. joining any two topological
images of a single object) is labelled $2\rinj,$ 
the {\em injectivity diameter}, while the diameter of the smallest
sphere, in the covering space, which contains the fundamental
polyhedron is termed $2r_+,$ the {\em out-diameter}.

It is possible that $\rinj \ll r_+$. For  
hyperbolic (negative curvature) 3-manifolds, of which an 
infinite number are topologically distinct, it is common to
have small values of $\rinj.$ This means it would be possible
to have many topological images in some directions of the observable
sphere, but no topological imaging in other directions.

In the candidate hyperbolic manifold of \citeauthor{BPS98} 
(\citeyear{BPS98}, \S4.3), although the volume of the space
is larger than that of the observable sphere, the 
injectivity diameter is slightly smaller, i.e. $\rinj \approx 0\.96 R_H.$
That is, 
if population III globular cluster-like objects 
were visible just after recombination,
then in some directions of the sky they would have topological images
visible at certain positions on the CMB, assuming a future CMB satellite
can reach the required resolution and sensitivity.

The candidate manifold of \citet{RE97} is more readily falsifiable:
$965\pm5$~{\hMpc} $< 2\rinj <$ $1190\pm10$~{\hMpc} depending on the 
values of the curvature parameters. For a review of arguments for
and against this candidate, see the discussion section of \citet{RBa99}.

\subsubsection{Curvature and distance-redshift relations} \label{s-curv}

In the covering space, relations between the redshift
and the proper distance (projected onto the present epoch, $t=t_0$),
depend on the curvature parameters indicated above: $\Omega_0$
and $\lambda_0.$

The proper distance to a redshift $z$ is
\begin{equation}
d(z) = {c\over H_0} \int_{1/(1+z)}^{1} 
{\mbox{\rm d}a \over a \sqrt{\Omega_0/a - \kappa_0 + \lambda_0\, a^2}} 
\label{e-dprop}
\end{equation}
where $\kappa_0 \equiv \Omega_0 + \lambda_0 -1$ as above.
This is related to the curvature radius 
\begin{equation}
R_C \equiv {c \over H_0 \sqrt{|\kappa_0|}} \label{e-defR_C}
\end{equation}
and the proper motion distance $\dpm(z)$ by
\begin{equation}
	d(z) = \left\{ 
	\begin{array}{lll}
	R_C \sinh^{-1} [\dpm(z)/R_C] , & \kappa_0 < 0 \\
	\dpm(z) , & \kappa_0 = 0 \\
	R_C \sin^{-1} [\dpm(z)/R_C] , & \kappa_0 > 0.
	\end{array}
	\right.
\label{e-d_dpm}
\end{equation}
If $\Omega_0 > 0$ and $\lambda_0=0,$ then 
\begin{equation}
\dpm(z) = {c \over H_0} 
    { 2 [ z\Omega_0 + (\Omega_0-2)(\sqrt{\Omega_0 z + 1} -1) ]
       \over \Omega_0^{\;2} (1+z) }   \label{e-dpm_0cc}
\end{equation}
is a commonly used closed expression for the proper motion distance 
(\citealt{Wein72}, p.485). For non-zero values of $\lambda_0,$
eqs~(\ref{e-dprop}), (\ref{e-defR_C}) and (\ref{e-d_dpm}) provide an
integral expression for $\dpm$. Values of these distances 
in physical units of typical astrophysical objects and of the
particle horizon are shown in Fig.~1 of \citet{RB98}.

\subsection{Uncertainties in $\Omega_0$ and $\lambda_0$} \label{s-uncert}

An order of magnitude estimate of the uncertainty in the curvature
parameters can be made as
follows.  
For an object seen at 
$z\sim 3,$ $\Delta z \sim 0\.002$ corresponds to a precision of $\sim
1${\hMpc}. 
The proper distance to $z=3$ ranges from $3000${\hMpc} 
to $\approx 5000${\hMpc} for 
$1 > \Omega_0 > 0\.2,$ $\lambda_0 \equiv 1-\Omega_0.$ 

Then the difference in proper distance over the full range of
curvature parameters is $\approx 2000${\hMpc}, so that a precision
of 1{\hMpc} corresponds to $0\.05\%.$ If the
dependence of proper distance on $\Omega_0$ and $\lambda_0$ 
were roughly linear over this range, then 
for a fixed 
value of the proper distance to the object at $z=3$,
the uncertainties
in  $\Omega_0$ and $\lambda_0$ would also be less than $0.1$\%. 
This is at least as precise as expected from the future 
CMB satellites Planck and MAP.

\fgeom 

Of course, this is only an order of magnitude estimate, which doesn't
take into consideration the actual numbers and relative geometry
of multiple topological images.
In reality, many topological images would have to be identified
before a candidate manifold could be considered significant, 
and the accuracy of the 
values of the curvature parameters would be iteratively 
improved as more topological images are included in a best fit
solution. 

\subsubsection{Separation into radial and tangential components}
\label{s-sepnradtang}

The objects whose three-dimensional positions 
are most sensitive to the curvature parameters are
those at the highest redshifts. Both the radial and tangential
components will be more sensitive at higher redshifts. If the
curvature is negative, then due to
the $\sinh$ factor the tangential components should be more
sensitive than the radial components, and vice versa for 
positive curvature.

The separations at lower
redshifts, which would be less sensitive 
to curvature parameters, could be considered as relatively fixed 
values for use at the higher redshifts.

To characterise these dependences, idealised cases of 
topological images separated in radial and tangential 
directions, shown in Fig.~\ref{f-geom}, are considered.
These are only fictional 
constructs, particularly the latter, 
since the value of $2\rinj$ must be a large fraction
($\gg 1\%$) of a horizon radius, so that $\theta_i \ll 1$~rad$, i=1,2$
is not valid, and the spatial geodesics 
joining the pairs of images at $z_1$ and $z_2$ (as opposed to the
arcs joining them) will not be equal, except in special cases.

Small variations in the redshifts, due to spectroscopic uncertainty 
or to the inability to correct for peculiar velocities, then imply
variations in $\Omega_0$ and $\lambda_0$ needed in order to retain
the equalities in the radial or tangential 
distances defining $z_0, z_1,$ $z_2,$ $\theta_1$ and $\theta_2.$
The resulting quantities are 
$\partial \Omega_0 /\partial z_i$ and 
$\partial \lambda_0 /\partial z_i$, where $i=0,1,2$ in the radial
direction and $i=1,2$ in the tangential direction.

In a real case, the separations between topological images will not
generally be either radial nor tangential, so the uncertainties 
will be somewhere in between these two limiting cases. 

This is equivalent to inverting eq.~(\ref{e-dprop}) 
so that the curvature parameters are 
functions of proper distance and redshift. 

\subsubsection{Cosmological redshift uncertainties}

To determine the uncertainties in $\Omega_0$ and $\lambda_0,$
the derivatives $\partial \Omega_0 /\partial z_i$ and 
$\partial \lambda_0 /\partial z_i$  need to be combined with
the uncertainty in the cosmological component of the redshift.

As pointed out in section 2 of \citet{Rouk96}, this uncertainty
in the cosmological (smooth expansion) component of the redshift
can be separated into 
(1) the spectroscopic 
uncertainty,\footnote{Photometric redshifts presently attain
a precision of $\sim 0.1$ (e.g. \protect\citealt{MPR99}), but this 
is too imprecise for topological purposes.} 
(2) two components due to the peculiar velocity:
(2a) movement of an object between two different epochs and
(2b) the error caused by assuming the observed redshift to
be purely cosmological. For physically reasonable values of 
the peculiar velocity and redshifts $z \ltapprox 3,$ the 
error (2a) is a small fraction of that due to (2b).

A typical precision practical for the spectroscopic uncertainty 
is $\Delta z \sim 0\.001$ for an object such as a quasar, though
much higher precision is possible. In that case the uncertainty
is dominated by the peculiar velocity. A conservative upper limit
to the peculiar velocity, assuming the quasar to be at the centre
of a typical galaxy, can be taken as $\Delta(zc) \sim 600$\k, 
i.e. $\Delta z \sim 0\.002.$

For a cluster of galaxies, spectroscopic redshifts could be found
very precisely for individual galaxies, but 
a central velocity would have to be obtained from a fit to the 
distribution of the galaxies' velocities, 
assuming isotropy of the peculiar velocities within the cluster. 
The X-ray velocity of the peak of the X-ray
distribution would provide another way to estimate the true
cluster redshift, inclusive of the peculiar velocity of the 
cluster as a whole.

If two topological images of the cluster were at similar redshifts, as
in the candidate identity suggested by \citet{RE97}, then as pointed
out by \citet{RBa99}, transverse velocities could be measured for the
galaxies in the cluster, and it would be possible to derive a detailed
dynamical model implying an even preciser cluster velocity.

However, the peculiar velocity of the cluster as a whole would
probably dominate the error. Here, $\Delta z \sim 0\.002$ 
is adopted as a typical value.

For `objects' such as cold spots in the CMB \citep{CaySm95},
the uncertainty in the redshift would be related to the thickness
of the SLS. 
The latter is $\Delta z \sim 100$ (and the SLS redshift 
is $z\sim 1100$; \citealt{WSS94,BondLH96}). This corresponds 
to a comoving thickness of $\sim 10-20{\hMpc}.$

\subsubsection{Uncertainties in angles}

Astrophysical objects at sub-horizon distances such as quasars
and clusters of galaxies can have their angular positions
measured much more precisely ($\sim 1\arcs$ for quasars; 
a few arcseconds for clusters if gravitational lensing is
available)
than their cosmological redshifts.
So, the error contributed to relative distance estimates is negligible
relative to the error from redshift uncertainty.

If identification of COBE cold spots with local 
superclusters were successful, then the poor angular 
resolution of COBE ($\sim 10\deg$ FWHM) 
would introduce an uncertainty of $\sim 15\%$ in 
the tangential components at the redshift of the SLS,
so the total uncertainties $\Delta\Omega_0$ and $\Delta\lambda_0$ 
would be much larger than just 
$\partial \Omega_0 /\partial z_i \Delta z_i$ and 
$\partial \lambda_0 /\partial z_i \Delta z_i.$


In contrast, Planck and MAP, with expected resolutions 
of $\sim 0\.1\deg$ and $\sim 0\.3\deg$ respectively,
would provide good tangential constraints.

\subsubsection{An illustrative simulation} \label{s-simmeth}

\fconfirm

The format according to which the first significant detections may
be made is unpredictable, but to illustrate the way 
the radial and tangential 
uncertainties in the curvature parameters combine, a simulation is
made in which it is hypothesized
that cluster images are observed at 
the positions and redshifts predicted by the 
candidate 3-manifold of \citet{RE97}.

According to this candidate 3-manifold, which can be 
called (with a slight abuse of language) 
a 2-torus, 
$T^2,$ (though really it is $T^2 \ttimes R$),
the size of the Universe is $2\rinj \approx (965\pm5)${\hMpc} for $\Omega=1$ 
[$2\rinj\approx (1190\pm10)${\hMpc} for $\Omega_0=0.2, \lambda_0=0.8$]
in two nearly perpendicular directions, 
and in the third perpendicular direction the size is unknown, 
i.e. $r_+ > \rinj$. 
Photons approaching the observer from greater distances 
in these directions would have already crossed the Universe once
or more, so the rich clusters RX~J1347.5-1145 and CL~09104+4109 
would be images of the Coma cluster seen about $2\.8${\hGyr}
ago for $\Omega=1$ ($3\.3${\hGyr} for 
$\Omega_0=0.2, \lambda_0=0.8$).

In the covering space, images of this cluster should form a nearly
square lattice (grid) out to the redshift at which the cluster formed.
For the simulation, it will be assumed that the first `confirmations'
of the hypothesis are detections of clusters at the `antipodal' directions
to RX~J1347.5-1145 and CL~09104+4109, 
(the positions are antipodal
from a Coma-centred viewpoint) and that successive confirmations 
of images are made first to successively higher redshifts, 
and then in roughly `tangential' directions. This is, of course, an
idealised simulation, ignoring systematic problems like foreground 
clusters or galaxies, dust, etc. This pattern of `confirmation' 
is shown in Fig.~\ref{f-confirm}.

The simulations are performed as follows. 
For a given choice of $\Omega_0$ ($\lambda_0$ is 
constrained as $\lambda_0= 1-\Omega_0$ since the model is flat),
$N=100$ simulated sets $\{ z_i\}_{i=4,\imax}$  are generated.
The $z_i$ values are calculated by assuming the generators to
be exactly the vectors from Coma to 
RX~J1347.5-1145 and CL~09104+4109, where spectroscopic and peculiar
velocity errors are assumed to be zero for these three images,
and by choosing a single combined spectroscopic plus peculiar velocity 
error for each $i=4,...,\imax$ from a gaussian distribution centred
at zero with dispersion $\sigma=0.002.$ 

For a given simulation, the `observational' deduction of the best
fit $\Omega_0$ value needs to be found. Over the $N$ simulations,
the distribution in the differences between input and output 
values of $\Omega_0$ determines the uncertainty in the value of
$\Omega_0.$

The method of finding the `best fit' value of $\Omega_0$ for 
a given simulation is to find the value of $\Omega_0$ for 
which the `images' of the two generators are most self-consistent.
That is, the value of $\Omega_0$ is found which minimises
\begin{eqnarray}
\label{e-defselfconsis}
\sigma^2 &=& \sigma_1^2 + \sigma_2^2 \nonumber \cr
&=& \sum_i^{\imaxtiny} d_1(i,j)^2
+ \sum_i^{\imaxtiny} d_2(i,j)^2
\end{eqnarray}
where the values of $(i,j)$ used for distances $d_1$ are
4--1, 1--3, 2--6, 3--7, 6--8, 7--9, 8--10, 9--11 and 10--12 
(up to the value of $\imax$) 
and those for $d_2$ are
1--2, 1--5, 3--6, 7--8, 9--10, 11--12
(up to the value of $\imax$). See Fig.~\ref{f-confirm} for 
the image numbering. This uses a close to maximal amount of nearly 
independent information, i.e. for what in principle are two
spikes in the crystallographic method \citep{LLL96}, but a single spike in
practice since the two generators are of equal lengths to within 1\%.


\section{Results} \label{s-res}

\fomBradi


\subsection{Radial and tangential components}

The relations $\partial \Omega_0 /\partial z_i$ and 
$\partial \lambda_0 /\partial z_i$ have been calculated for
values of the curvature parameters spanning the values of
observational interest, and multiplied by $\Delta z = 0\.002$
to give estimates of $\Delta\Omega_0$ and $\Delta\lambda_0.$ 
These are shown in Figs~\ref{f-om_1000radi} and \ref{f-om_1000tang}.

As expected, the figures clearly show that 
\begin{list}{(\roman{enumi})}{\usecounter{enumi}}
\item the higher redshift objects
provide preciser constraints than the low redshift objects
\item the tangential constraints are preciser than the radial ones
\item for sets of objects at redshifts $1+z < 4,$ a precision
of better than 1\% in $\Omega_0$ and 10\% in $\lambda_0$ is possible.
\end{list}

Any single set of three topological images separated in a {\em radial
direction} would correspond to a combination of three points 
on the curves of a single line style 
in Fig.~\ref{f-om_1000radi}.
Assuming uncorrelated gaussian errors in the three redshifts $z_i$
implies total uncertainties of 
\begin{eqnarray}
\Delta\Omega_0 
&=&\sqrt{\sum_i (\partial \Omega_0 /\partial z_i \; \Delta z_i)^2} 
\nonumber \\
\Delta\lambda_0 
&=&\sqrt{ \sum_i (\partial \lambda_0 /\partial z_i \; \Delta z_i)^2}
\label{e-domtot}
\end{eqnarray}
where $i=0,1,2.$

Any set of four (idealised) topological images forming 
{\em tangentially separated} pairs
at $z_1$ and $z_2$ as in Fig.~\ref{f-geom} would similarly 
correspond to a combination of two points on the curves 
of a single line style in 
Fig.~\ref{f-om_1000tang}.
Equation (\ref{e-domtot}) would then give the total uncertainties,
where $i=1,2.$


\fomAtang


A real set of topological images would have a less simple set
of orientations, so the real uncertainties for a set of three or
four objects
would consist of interpolations between these two components.

The limiting term in improving the precision of the constraints
is clearly the lowest redshift of the triplet or the quadruplet.

Note that $\partial \lambda_0 /\partial z_i$ changes sign at
a low redshift (the plots show absolute values, so this appears
as a cusp). This is simply due to the transition between the
regimes where $\lambda_0$ has a negligible effect on the metric
and where it has a significant effect.
The dependence of $\lambda_0$ on changes in $z_i$ 
changes sign at low redshifts (the plots show absolute values),
below which the effect of $\lambda_0$ on the radial component
of the metric becomes weaker than that of $\Omega_0.$

It should also be noted that the precision of $10^{-6}$ to
$10^{-7}$ for images at $z \sim 1000$ 
(off the scale of the figures as shown) 
is not of practical
significance for known astrophysical objects, though 
if a generation of $\sim 10^6 M_\sun$ collapsed objects 
containing population III stars existed following 
the recombination epoch, their eventual detection is not
totally impossible.

The precision of $10^{-6}$ to $10^{-7}$ at the SLS itself
would only be valid if sub-SLS objects were identified with
CMB features, and if the redshift of the features in the
CMB were precise to $\Delta z = 0\.002.$ If a more realistic
figure of $\Delta z \sim 100$ (as mentioned above) is used 
for the CMB `spots', then the uncertainty in $\Omega_0$
and $\lambda_0$ is a factor of $5\e{4}$ times higher than
shown in the figures, i.e. the precision is about the
same order of magnitude as that possible from sub-SLS objects.

\subsection{Simulations}

\tsim

Simulations as described in \S\ref{s-simmeth} were performed
for the popular values of $\Omega_0=0.3, \lambda_0=1-\Omega_0$,
supposing that successive observations reveal cluster images at
the positions predicted from the \citet{RE97} $T^2 \ttimes R$ candidate
in the order illustrated in Fig.~\ref{f-confirm}.
Since the density parameter is low in this model, the birth
of the cluster can be at a quite high redshift (though this also
depends on other parameters like the slope of the primordial
power spectrum).

Table~\ref{t-sim} shows the resulting uncertainties in the
value of $\Omega_0.$ As expected, accuracies greater than 1\%
can easily be obtained from a dozen images 
with $z \ltapprox 3.$

\section{Discussion} \label{s-disc}

The calculations presented show the precision obtained 
if only a few topological images have been reliably identified.
However, once this has been done, further multiple topological 
images of the same object and multiple topological images of
other objects will be easier to find. This should yield preciser
values still of both the curvature parameters and the generators,
so the finding of new corresponding topological images would 
increase exponentially until saturated by the limits of 
observational catalogues.

The improvement in accuracy can be estimated as follows.

We can consider a set of four objects at relative positions,
neither aligned radially nor tangentially, as providing {\em both} the
radial and tangential separation components as presented above. 
In that case, if there are a total of $N$ multiple topological images
of a single object (e.g. the brightest X-ray cluster known), these
can be considered as $N/4$ independent sets of four images.
Label the maximum uncertainties for any such set of four images
as $\Delta(\Omega_0,4)$ and $\Delta(\lambda_0,4).$

Label the number of physically distinct objects, for which each 
has about $N$ multiple topological images identified, as $M.$

Then, the independence of gaussian errors implies a reduction 
in the uncertainties:  
\begin{eqnarray}
\Delta(\Omega_0,MN) &\sim& \Delta(\Omega_0,4)/ \sqrt{MN/4} \nonumber \\ 
\Delta(\lambda_0,MN) &\sim& \Delta(\lambda_0,4)/ \sqrt{MN/4}. 
\end{eqnarray}

The total number of galaxies visible
to apparent magnitude limits of $V \sim 26$ is 
$MN \sim 10^{10}$--$10^{11}.$ 
So, if $2\rinj$ is small enough that the majority of galaxies have 
at least four multiple images at epochs later than the galaxies'
formation epoch, then the ultimate precision in estimating
the curvature parameters could reach around $10^{-6}$--$10^{-7}$,
once all-sky redshift surveys with spectroscopy to a precision
better than $\Delta z \sim 0\.001$ and complete to $V \sim 26$ 
are performed. This is not attainable in the coming decade, 
but could be envisaged as a project for 
the New Generation Space Telescope (NGST).

\section{Conclusions}\label{s-conc}

If our Universe corresponds to a 
Friedmann-Lema\^{\i}tre model, and if the hypothesis of 
a trivial topology is observably wrong, then
the detection of multiple topological images would
enable constraints on the curvature parameters from
known astrophysical objects to be made as precisely as those
presently expected for the Planck and MAP satellites.

The dependence of the precision on the redshifts
of the multiple topological images and on their radial
and tangential separations has been presented and
calculated. 

The tangential separations give tighter
constraints. Sets of multiple topological images 
at redshifts $z \ltapprox 3$
would imply values of $\Omega_0$ and $\lambda_0$ 
preciser than $\sim 1\%$ and $\sim 10\%$ respectively.
The precision available from the Planck and MAP satellites 
for CMB `objects' cross-identified with 
low redshift objects would be similar.

Looking further into the future, an all sky spectroscopic survey
by the NGST could lead to precision on a scale impressive
by today's standards: $10^{-6}$--$10^{-7}$. Observational 
cosmology would shift from the phase of working out the basics to 
that of high precision science.


\end{document}